\documentclass[iop]{emulateapj}
\usepackage{apjfonts}
\usepackage{graphicx}
\usepackage{mediabb}

\newcommand{\kanata}{\textit{Kanata}}
\newcommand{\pirka}{\textit{Pirka}}
\newcommand{\swift}{\textit{Swift}}
\newcommand{\maxi}{\textit{MAXI}}
\newcommand{\fermi}{\textit{Fermi}}
\newcommand{\integral}{\textit{INTEGRAL}}

\newcommand{\src}{V404~Cyg}

\newcommand{\erg}{\mathrm{erg}}

\newcommand{\s}{\mathrm{s}}

\def\deg{\hbox{$^\circ$}}

\slugcomment{Accepted by ApJ}

\begin{document}

\title{No evidence of intrinsic optical/near-infrared linear polarization for V404 Cygni during its bright outburst in 2015: Broadband modeling and constraint on jet parameters}

\author{Y.~T.~Tanaka\altaffilmark{1}, R.~Itoh\altaffilmark{2}, M.~Uemura\altaffilmark{1}, Y.~Inoue\altaffilmark{3}, C.~C.~Cheung\altaffilmark{4}, 
M.~Watanabe\altaffilmark{5}, K.~S.~Kawabata\altaffilmark{1}, Y.~Fukazawa\altaffilmark{1}, 
Y.~Yatsu\altaffilmark{6}, T.~Yoshii\altaffilmark{6}, Y.~Tachibana\altaffilmark{6}, T.~Fujiwara\altaffilmark{6}, Y.~Saito,\altaffilmark{6}, N.~Kawai\altaffilmark{6}, 
M.~Kimura\altaffilmark{7}, K.~Isogai\altaffilmark{7}, T.~Kato\altaffilmark{7}, 
H.~Akitaya\altaffilmark{1}, M.~Kawabata\altaffilmark{2}, T.~Nakaoka\altaffilmark{2}, K.~Shiki\altaffilmark{2}, K.~Takaki\altaffilmark{2}, M.~Yoshida\altaffilmark{1},
M.~Imai\altaffilmark{5}, S.~Gouda\altaffilmark{5}, Y.~Gouda\altaffilmark{5}, 
H.~Akimoto\altaffilmark{8}, S.~Honda\altaffilmark{8}, K.~Hosoya\altaffilmark{8}, A.~Ikebe\altaffilmark{8}, K.~Morihana\altaffilmark{8}, T.~Ohshima\altaffilmark{8}, 
Y.~Takagi\altaffilmark{8}, J.~Takahashi\altaffilmark{8}, K.~Watanabe\altaffilmark{8},
D.~Kuroda\altaffilmark{9}, T.~Morokuma\altaffilmark{10}, K.~Murata\altaffilmark{11}, T.~Nagayama\altaffilmark{12}, D.~Nogami\altaffilmark{7}, Y.~Oasa\altaffilmark{13}, 
K.~Sekiguchi\altaffilmark{14}
}

\email{ytanaka@hep01.hepl.hiroshima-u.ac.jp}

\altaffiltext{1}{Hiroshima Astrophysical Science Center, Hiroshima University, 1-3-1 Kagamiyama, Higashi-Hiroshima 739-8526, Japan}
\altaffiltext{2}{Department of Physical Sciences, Hiroshima University, Higashi-Hiroshima, Hiroshima 739-8526, Japan}
\altaffiltext{3}{Institute of Space and Astronautical Science, JAXA, 3-1-1 Yoshinodai, Chuo-ku, Sagamihara, Kanagawa 252-5210, Japan}
\altaffiltext{4}{Space Science Division, Naval Research Laboratory, Washington, DC 20375-5352, USA}
\altaffiltext{5}{Department of Cosmosciences, Graduate School of Science, Hokkaido University, kita 10, Nishi 8, Kita-ku, Sapporo, Hokkaido 060-0810, Japan}
\altaffiltext{6}{Department of Physics, Tokyo Institute of Technology, 2-12-1, Ohokayama, Tokyo, Japan}
\altaffiltext{7}{Department of Astronomy, Kyoto University, Kitashirakawa Oiwake-cho, Sakyo-ku, Kyoto, Kyoto 606-8502, Japan}
\altaffiltext{8}{Center for Astronomy, University of Hyogo, 407-2 Nishigaichi, Sayo, Hyogo 679-5313, Japan}
\altaffiltext{9}{Okayama Astrophysical Observatory, National Astronomical Observatory of Japan, Asakuchi, Okayama 719-0232, Japan}
\altaffiltext{10}{Institute of Astronomy, Graduate School of Science, The University of Tokyo, Mitaka, Tokyo 181-0015, Japan}
\altaffiltext{11}{Graduate School of Science, Nagoya University, Furo-cho, Chikusa-ku, Nagoya 464-8602, Japan}
\altaffiltext{12}{Graduate School of Science and Engineering, Kagoshima University, Kagoshima 890-0065,Japan}
\altaffiltext{13}{Faculty of Education, Saitama University, Sakura, Saitama 338-8570, Japan}
\altaffiltext{14}{National Astronomical Observatory of Japan, Mitaka, Tokyo 181-8588, Japan}

\begin{abstract}
We present simultaneous optical and near-infrared (NIR) polarimetric results for the black hole binary \src\ spanning the duration of its 7-day long optically-brightest phase of its 2015 June outburst. The simultaneous $R$ and $K_s$-band light curves showed almost the same temporal variation except for the isolated ($\sim 30$ min duration) orphan $K_s$-band flare observed at MJD~57193.54. We did not find any significant temporal variation of polarization degree (PD) and position angle (PA) in both $R$ and $K_s$ bands throughout our observations, including the duration of the orphan NIR flare. We show that the observed PD and PA are predominantly interstellar in origin by comparing the \src\ polarimetric results with those of the surrounding sources within the $7' \times 7'$ field-of-view. The low intrinsic PD (less than a few percent) implies that the optical and NIR emissions are dominated by either disk or optically-thick synchrotron emission, or both. We also present the broadband spectra of \src\ during the orphan NIR flare and a relatively faint and steady state by including quasi-simultaneous \swift/XRT and \integral\ fluxes. By adopting a single-zone synchrotron plus inverse-Compton model as widely used in modeling of blazars, we constrained the parameters of a putative jet. Because the jet synchrotron component cannot exceed the \swift/XRT disk/corona flux, the cutoff Lorentz factor in the electron energy distribution is constrained to be $<10^2$, suggesting particle acceleration is less efficient in this microquasar jet outburst compared to AGN jets. We also suggest that the loading of the baryon component inside the jet is inevitable based on energetic arguments.
\end{abstract}


\clearpage

\section{Introduction}
\label{sec-intro}
A relativistic collimated outflow (known as a jet) emerging from a black hole (BH) is ubiquitously observed in various spatial scales from stellar mass ($\sim 10 M_{\odot}$) to supermassive ($\sim 10^8-10^{10} M_{\odot}$) BHs. Polarimetry in the optical and near-infrared (NIR) bands is a powerful method to unveil the emission mechanism and investigate the magnetic field structure inside the jet \citep[e.g.,][]{Marscher08}. In Active Galactic Nuclei (AGN) and Gamma-Ray Burst (GRB) jets, it is well accepted that the optical and NIR lights are produced by high-energy electrons through optically-thin synchrotron emission. High degrees of linear polarization in the optical/NIR band observed from AGN and GRB jets \citep[$\sim 10$\% up to 30--40\%; see e.g.,][]{Ikejiri11, Uehara12} indicate that the synchrotron process is operating, with highly ordered magnetic fields in the emission regions. Measurements of polarization position angle (PA) are also useful to determine the magnetic field direction at the emission region. In addition, the detection of a polarization PA swing may manifest from the presence of a helical magnetic field along a jet or curved structure indicating the global jet geometry \citep[e.g.][]{Marscher08, Abdo10}.

It is also known that the jet emission in stellar-mass BHs appear in the NIR band. While emission in the optical-band is dominated by the accretion disk, an excess with respect to the Rayleigh-Jeans tail of the disk blackbody component is often found in the NIR-band. Hence, if the NIR emission is due to \emph{optically-thin} synchrotron emission, a high polarization degree (PD) is theoretically expected. However, reliable optical/NIR polarimetric measurements for Galactic BH binary jets are still very limited mainly due to the difficulty in eliminating the interstellar polarization caused by large amounts of dust clouds in our Galaxy in the target directions.
Another reason for the paucity of intrinsic polarization measurements is that the objects are often not sufficiently bright to perform polarimetry. In this regard, we note that clear NIR polarization was detected from Cyg X-2 and Sco X-1 based on spectro-polarimetry \citep{Shahbaz08}.

An opportunity to study the BH binary jet through optical/NIR polarimetry was presented when V404 Cygni (a.k.a., GS~2023+338; hereafter \src) produced an exceptionally bright outburst in June 2015. This object is one of the famous low-mass X-ray binaries (LMXBs) because a similar huge outburst was detected in 1989 with intensive multi-wavelength observations performed at that time \citep[e.g.,][]{Makino89}. The distance is accurately determined as $2.39 \pm 0.14$~kpc from parallax measurement using astrometric VLBI observations \citep{Miller-Jones09}. In this paper, aimed at studying the non-thermal jet emission and constraining the physical parameters in a microquasar jet, we present results of linear polarization measurements in the optical and NIR bands for \src\ during the brightest outburst in June 2015 performed by the \kanata\ 1.5~m and \pirka\ 1.6~m telescopes in Japan. Observations and data reductions are described in \S\ref{sec-obs}. We show the results in \S\ref{sec-results}, and the implications of our findings are presented in \S\ref{sec-dis}.

\section{Observations and Data Reductions}
\label{sec-obs}

\subsection{\kanata/HONIR}
\label{sec-kanata}
We performed simultaneous optical and NIR imaging polarimetry for \src\ using the Hiroshima Optical and Near IR camera \citep[HONIR;][]{Akitaya14} mounted on the \kanata\ 1.5-m telescope in Higashi-Hiroshima, Japan. The data presented here were taken on MJD~57193 and 57194. The HONIR polarization measurements utilize a rotatable half-wave plate and a Wollaston prism. We selected $R_C$ and $K_s$ bands as the two simultaneous observing filters. To study the wavelength dependence of polarization properties for the target, we also took $V R_C I_C J H K_s$ imaging polarimetric data on MJD~57194.50--57194.55 (the ``C'' subscripts are herein suppressed). Each observation consisted of a set of four exposures at half-wave plate position angles of 0\fdg0, 22\fdg5, 45\fdg0, and 67\fdg5. Typical exposures in each frame were 30~s and 15~s for the $R$- and $K_s$-bands, respectively, but these exposures were sometimes modified depending on weather conditions (decreased when seeing became good, and increased when cirrus clouds passed over the target). $V I$- and $JH$-band exposures on MJD~57194 were 30~s and 15~s, respectively. 

To calibrate these data, we observed standard stars on MJD~57195 that are known to be unpolarized (HD~154892) and strongly-polarized \citep[HD~154445 and HD~155197;][]{Tur90, Wol96}. We thereby confirmed that the instrumental PD is less than 0.2\% and determined the instrumental polarization PA against the celestial coordinate grid.
Absolute flux calibration, which is needed to construct the optical and NIR SEDs, was performed by observing standard stars on the photometric night MJD~57200.

\subsection{\pirka/MSI}
\label{sec-pirka}
We also performed optical $R$-band imaging polarimetry monitoring for \src\ using the Multi-Spectral Imager (MSI; Watanabe et al. 2012) mounted on the 1.6-m Pirka telescope located in Hokkaido, Japan, from MJD~57190--57193. The MSI observations were performed in a similar manner to the \kanata/HONIR ones, namely the MSI utilizes a rotatable half-wave plate and a Wollaston prism, and a series of four exposures were taken for each polarization measurement. The typical exposure time of each frame was 15~s. To remove the instrumental polarization ($p = 0.78$\%) and to calibrate the polarization PA, we used past MSI data of the two unpolarized stars \citep[BD+32 3739 and HD~212311;][]{Schmidt92} and three strongly-polarized stars \citep[HD~154445, HD~155197, and HD~204827;][]{Schmidt92} obtained on MJD~57167 and 57169. We also confirmed the polarization efficiency of $\sim 99.7$\% using a polarizer and flat-field lamp.

\subsection{\swift/XRT}
\label{sec-swift}
We analyzed X-ray data for \src\ taken with XRT onboard the \swift\ satellite using HEASoft version 6.16. The \swift/XRT data analyzed here were taken on MJD~57193 and 57194 (observation IDs: 00031403040 and 00031403046), which were almost simultaneous (within less than 1 hour) with the \kanata/HONIR multi-band photo-polarimetry data. Clean events of grade 0--12 within the source rectangle region (because the observation was in window-timing mode) were selected. After subtracting background counts selected from both sides of the source with rectangle shapes, the 0.5--10~keV events were utilized for spectroscopy. We generated ancillary response files with the \texttt{xrtmkarf} tool. Using XSPEC version 12.8.2, we roughly fit the data by assuming a disk blackbody plus power-law model, both modulated by Galactic absorption (i.e., \texttt{wabs*(diskbb+pow})), and converted the deabsorbed spectra to $\nu F_{\nu}$ fluxes.

\begin{figure*}[!th]
\begin{center}
\plotone{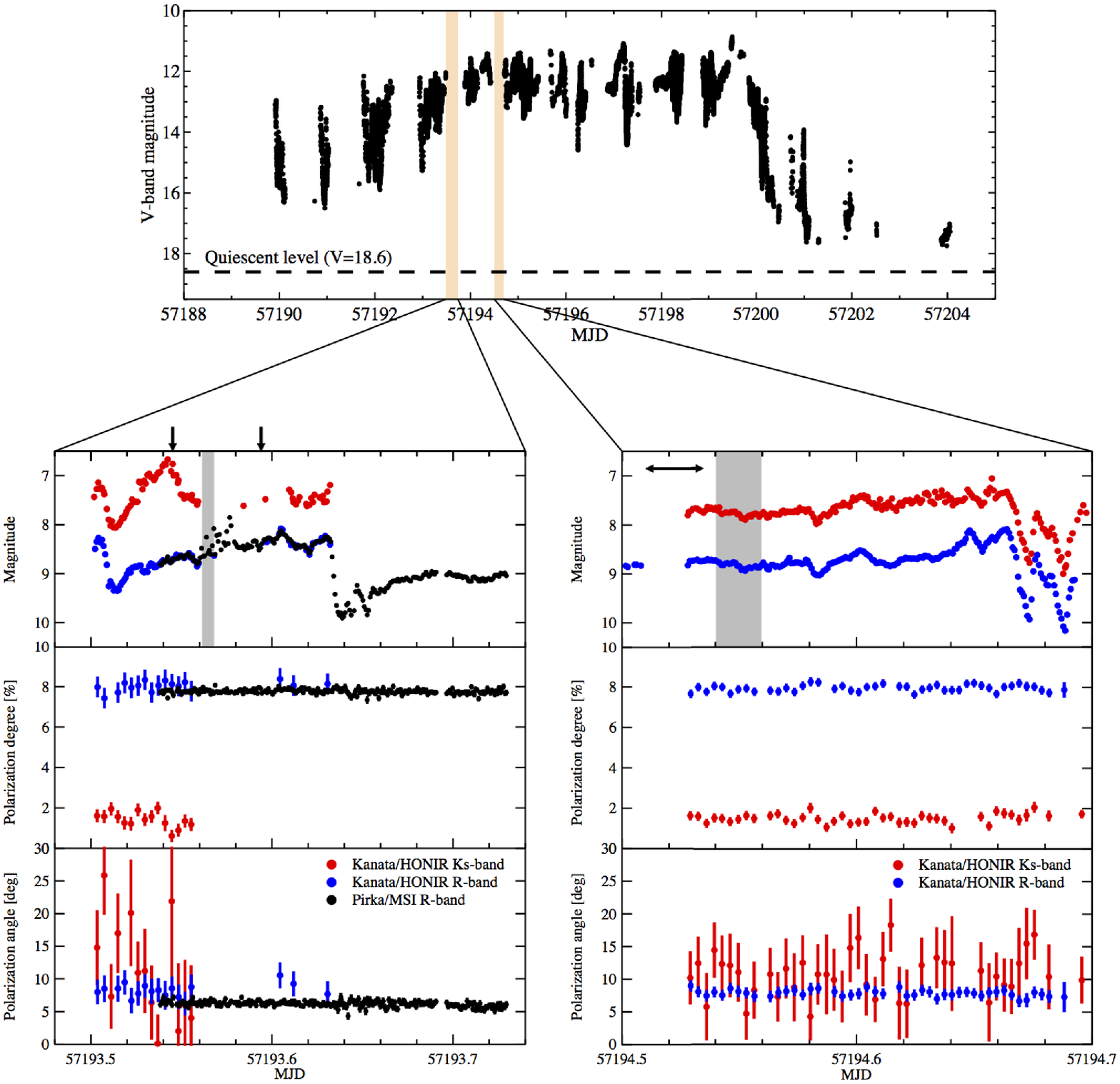}
\caption{{\it Top}: $V$-band light curve of \src\ during the most active phase in 2015 June \citep[taken from][]{Kimura16}. Horizontal dashed line indicates the quiescent flux level of $V=18.6$ \citep{Wagner91}. The hatched orange rectangles indicate the \kanata/HONIR observation periods shown in detail in the bottom multi-panel plots. 
{\it Bottom left panels}: \kanata/HONIR and \pirka/MSI polarimetric observations in MJD~57193. Note that the $R$-band magnitudes shown in the first panel are offset by 3.0 mag for illustrative purposes (i.e., the actual $R$-band magnitudes are 3.0 mag fainter). The two black arrows and gray rectangle in the first panel indicate the time ranges when \kanata/HONIR $VRIJHK_s$ and \swift/XRT X-ray spectra were constructed, respectively (see also {\it left} panels in Fig.~\ref{fig:fnu} and Fig.~\ref{fig:nufnu}). {\it Bottom right panels}: Same as {\it bottom left panels}, but for the data in MJD~57194. A black horizontal arrow indicates the period when Kanata/HONIR multiband polarimetric observations were performed (see also Fig.~\ref{fig:multiband} as well as {\it right} panels in Fig.~\ref{fig:fnu} and Fig.~\ref{fig:nufnu}).
}
\label{fig:lc}
\end{center}
\end{figure*}

\section{Results}
\label{sec-results}
Fig.~\ref{fig:lc} ({\it top} panel) shows the $V$-band light curve of \src\ during the bright outburst \citep{Kimura16} after the detection of burst-like activities by \swift/BAT, \fermi/GBM, and \maxi/GSC on 2015 June 15 (MJD~57188) \citep{MAXI_ATel, Swift_ATel}. The optical flux increased by $\sim 7$~mag compared to the quiescent level \citep[$V=18.6$~mag,][]{Wagner91} with a maximum around MJD~57194 and the highest flux level continued for about one week. During this brightest phase, the source showed large-amplitude (as much as 3~mag) and short-time variability. Fig.~\ref{fig:lc} (two {\it bottom} panels) illustrate intra-night variations of the $R$- and $K_s$-band fluxes, polarization degrees (PDs), and polarization position angles (PAs) measured by \kanata/HONIR and \pirka/MSI on MJD 57193 and 57194 (corresponding to 2015 June 19 and 20). Note that the \kanata/HONIR $R$- and $K_s$-band photometric and polarimetric observations are strictly simultaneous. On MJD~57193, the HONIR observations were interrupted by cloudy weather and stopped around MJD~57193.64, while \pirka/MSI continuously obtained $R$-band photometric and polarimetric data over $\sim 5$~hours. 

On the whole, the simultaneous $R$- and $K_s$-band light curves showed similar temporal variations. However, around MJD~57193.54, a flux increase is evident only in the $K_s$ band, while no corresponding enhancement was observed in the $R$-band. During this NIR flare, the $K_s$-band PD and PA did not show any significant variation despite the pronounced flux change. The $K_s$-band PD and PA were constant at $1.4\pm0.1$\% and $9\fdg1\pm2\fdg2$, respectively, throughout the HONIR observations that night. That same night, the $R$-band light curve showed a rapid and large-amplitude decrease around MJD~57193.64 and then gradually recovered. During the optical dip, the PD and PA were constant and did not show any significant variations. Similarly, the $R$-band PDs and PAs measured by \pirka/MSI on this night remained constant at $7.77\pm0.01$\% and $6\fdg19\pm0\fdg03$, respectively. Note that there is a small discrepancy between the \pirka/MSI and \kanata/HONIR $R$-band polarimetric results (see Table~\ref{tab:pol}). This could be due to a lack of cross-calibration but our subsequent discussion is unaffected by this small difference.

On the next night, the observing conditions were relatively good until MJD~57194.7 and we obtained $\sim 4.5$-hours of continuous, simultaneous $R$- and $K_s$-band photometric and polarimetric data for \src. As shown in Fig.~\ref{fig:lc} ({\it bottom-right} panel), the HONIR photometric light curves in $R$- and $K_s$ bands exhibited quite similar temporal profiles, including the two `dips' around MJD~57194.68. The PDs and PAs in both bands were again constant over the $\sim 4.5$-hour duration, even over the course of two observed flux dips. In addition, the PDs and PAs in each band were unchanged from those measured on the previous night (MJD~57193). We note the sporadic nature of the $K_s$-band polarimetric data points (namely, PD and PA) after MJD~57194.64 were due to the passage of cirrus clouds (NIR observations are more heavily affected by clouds compared to the optical).

Finally, we note that in addition to the photo-polarimetric data presented here we also obtained \pirka/MSI $R$-band data on MJD~57190, 57191, and 57192. As shown in Table~\ref{tab:pol}, these data showed that the polarization parameters of the object remained constant at  ${\rm PD}=7.7-7.9\%$ and ${\rm PA}=6\fdg2-6\fdg6$ over the three nights despite dramatic variability of the total flux of $\sim 2.0$~mag.

\begin{figure}[!th]
\begin{center}
\plotone{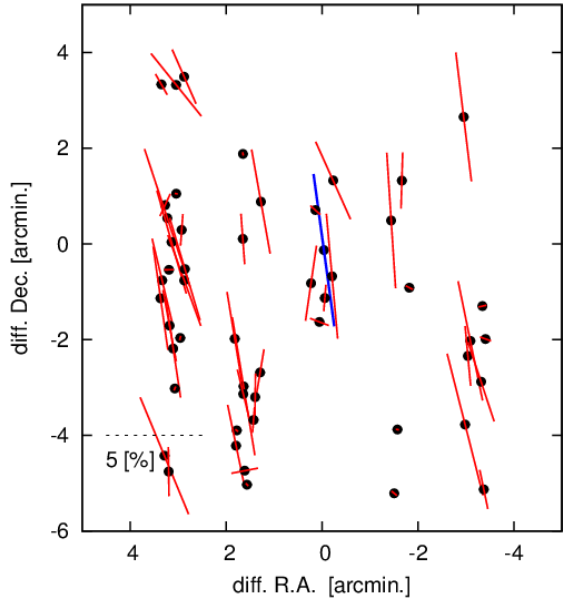}
\caption{$R$-band PD and PA for \src\ (shown in blue at center) and for surrounding sources (shown in red) within the $7' \times 7'$ \kanata/HONIR field-of-view.}
\label{fig:syuui}
\end{center}
\end{figure}

\begin{figure}[!th]
\begin{center}
\plotone{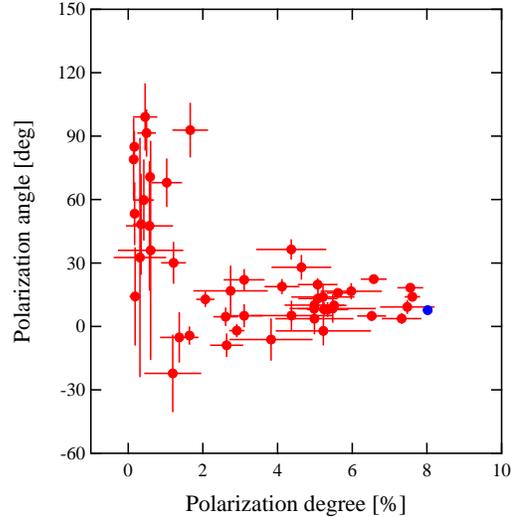}
\caption{$R$-band PD and PA for \src\ (shown in blue) compared to those observed from each source within the \kanata/HONIR field-of-view (shown in red); cf., Fig.~\ref{fig:syuui}.}
\label{fig:versus}
\end{center}
\end{figure}

To investigate the polarization properties of the sky region in the direction of \src, we also analyzed the \kanata/HONIR $R$-band data taken on MJD~57194 (selected because the observing condition was much better compared to MJD~57193) and determined the PDs and PAs for the brightest field stars within the HONIR field-of-view (FoV). The results are displayed in Fig.~\ref{fig:syuui}. We found that the PAs of not only \src, but also the surrounding objects, showed almost the same direction. Moreover, the measured PDs were also observed at similar levels. These findings clearly indicate that, despite the relatively large PD of $\sim 7.8$\% for \src, local dust clouds located between \src\ and the Earth are the likely cause of the polarized emission in this sky direction and about half of the surrounding objects (including \src) are located beyond the dust cloud. This suggests the observed PD and PA for \src\ is \emph{not} intrinsic \emph{but} interstellar origin. 
The hypothesis is supported by the non-variable PDs and PAs for \src\ observed even during the large flux variations (see two bottom panels in Fig.~\ref{fig:lc}). We also plot in Figure~\ref{fig:versus} the measured PA as a function of PD for each object within the FoV. Two clusterings of the data are clearly visible: objects with very small PD and a wide PA range over 180\deg\ are likely located in front of local dust clouds, while those with relatively large PDs of typically $\sim 5\%$ and broadly similar PAs of 0\deg--30\deg\ are positioned beyond the dust clouds. We note that the slightly greater PD of \src\ ($\sim 8\%$) with respect to the surrounding objects ($\sim 5\%$) implies a small level of intrinsic polarization for \src\ of at most a few percent.

\begin{figure}[!th]
\begin{center}
\plotone{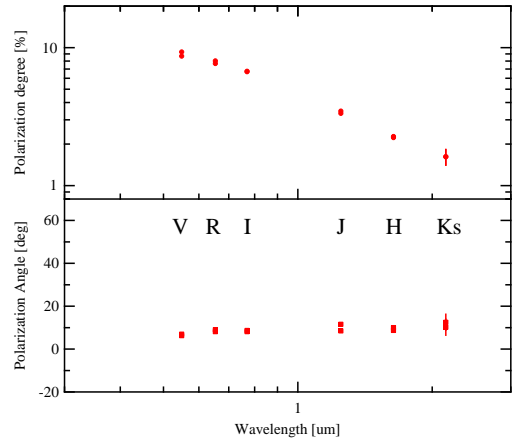}
\caption{Nearly simultaneous $VRIJHK_s$-band PD and PA measurements of \src\ observed with HONIR from MJD~57194.51--57194.54.
}
\label{fig:multiband}
\end{center}
\end{figure}

Furthermore, we show the $VRIJHK_s$-band PDs and PAs of \src\ in Fig.~\ref{fig:multiband} indicating a steep PD decrease toward longer wavelengths with constant PAs over the six observation bands. This polarization behavior is similar to that of a highly reddened star, suggesting that the polarization is interstellar origin. 
Detailed study of the interstellar dust based on these multi band polarimetric data will be reported in a forthcoming paper (Itoh et al. in preparation). From these observational results, we consider the measured polarization of \src\ is predominantly contaminated by interstellar dust between the object and the Earth. The low intrinsic PD (less than a few percent) implies that the optical and NIR emissions are dominated by either disk or optically-thick synchrotron emission, or both.

\begin{figure*}[!th] 
\begin{center}
\plotone{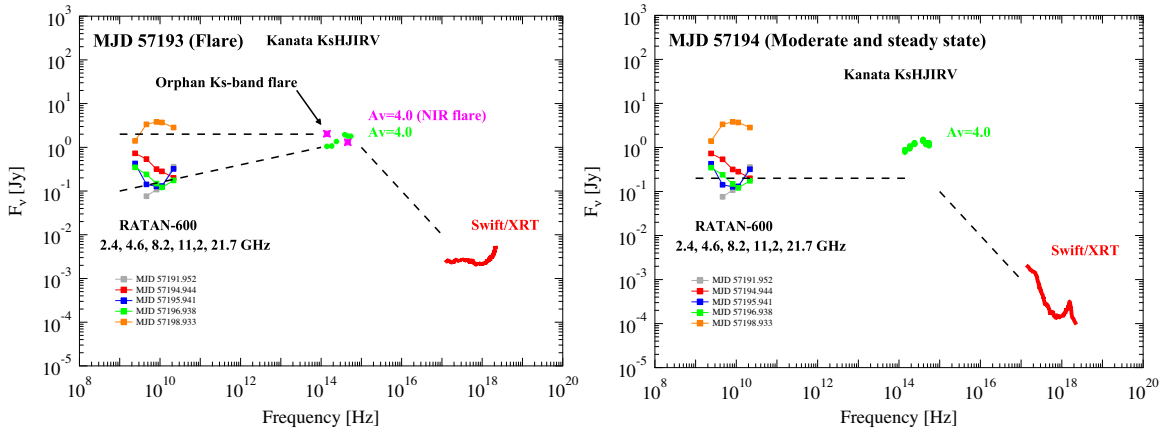}
\caption{{\it Left}: Quasi-simultaneous (within 1~hour) $\nu - F_{\nu}$ plotted measurements of \src\ observed by \kanata/HONIR and \swift/XRT on MJD~57193. The simultaneous $K_s$- and $R$-band fluxes measured during the orphan $K_s$-band flare are shown with magenta stars. 
{\it Right}: Same as the {\it left} panel, but for the \kanata/HONIR and \swift/XRT data taken on MJD~57194.
Also shown in both panels are preliminary non-simultaneous 2.3 to 21.7 GHz RATAN-600 fluxes from \citet{RATAN_ATel}. For illustrative purposes only, dashed lines indicating various spectral indices ($\alpha=0$ and $-1.0$ in both, plus additional $\alpha = 0.2$ in left panel; $F_{\nu} \propto \nu^{\alpha}$) are shown.
}
\label{fig:fnu}
\end{center}
\end{figure*}

\section{Discussion}
\label{sec-dis}
\subsection{Broadband spectrum of \src}
The simultaneous \kanata/HONIR $R$- and $K_s$-band light curves showed almost the same temporal evolution, except the orphan $K_s$-band flare which peaked around MJD~57193.54 and lasted for $\sim 30$~mins (see Fig.~\ref{fig:lc}). Apart from this orphan flare (which is discussed in detail later), the quite similar $R$- and $K_s$-band  light curves naturally leads to an interpretation that the NIR and optical emissions come from the same component (or have the same origin). There are two possible options to explain the optical and NIR emissions: one is a disk origin and the other is from a jet. To gauge which is the most plausible, we constructed a broadband spectrum of \src\ from the radio to X-ray bands in $\nu$-$F_{\nu}$ representation (Fig.\ref{fig:fnu}). Note that the radio fluxes are not simultaneous, while the \kanata\ and \swift/XRT data were obtained within a one-hour timespan. The \kanata\ optical and NIR fluxes were dereddened by assuming $A_V=4.0$ and $R_V=3.1$. The $A_V$ value of 4.0 was derived by \citet{Casares93} from the spectral type of companion star and $B-V$ colors, which was also confirmed by subsequent studies \citep[e.g.,][]{Shahbaz03, Hynes09}.
After we corrected the observed fluxes for extinction using $A_V=4.0$, we found a slightly rising, but almost flat ($\alpha \sim 0$) shape in the optical and NIR spectrum. This implies optically-thick synchrotron emission from an outer jet. Indeed, the $\sim 2-20$~GHz radio spectrum obtained about 1.5~days before our observation (on MJD~57191.95) can be smoothly extrapolated to the  \kanata/HONIR spectral data assuming a $F_{\nu} \propto \nu^{0.2}$ form (see Fig~\ref{fig:fnu}).
However, the flat optical/NIR spectral shape can also be interpreted in a disk model \citep[see e.g.,][Extended Data Figure~6 therein]{Kimura16}. Thus, it is difficult to determine the optical/NIR emission mechanism solely from its spectral shape. No evidence of intrinsic linear polarization in the $R$- and $K_s$-bands are allowed in both scenarios because both optically-thick synchrotron and blackbody radiation only generate weak linear polarization of order $\lesssim 10$\%. 

Here, we focus on the orphan $K_s$-band flare which lasted for only $\sim 30$ mins at MJD~57193.54. The observed red color and short duration imply synchrotron emission from a jet as the most plausible origin of the flare. Indeed, the $K_s$-band peak flux of the flare reached $\sim 2.0$~Jy, which was the same level measured during the giant radio and sub-mm flares observed by RATAN-600 and Sub Millimeter Array on MJD~57198.933 and MJD~57195.55 \citep{Trushkin7716, Tetarenko15}, respectively. As shown in Fig.~\ref{fig:fnu} ({\it left} panel), an extrapolation of the radio spectrum observed during the giant flare at GHz-frequencies on MJD~57198.933 nicely connects to the $K_s$-band peak flux by assuming a flat spectral shape (i.e., $F_{\nu} \propto \nu^{0}$). More interestingly, even during the orphan flare, the $K_s$-band PD remains showed no significant temporal variation, indicating the NIR emission is not strongly polarized. This result would be reasonably understood if the jet synchrotron emission in the $K_s$ band is still in the optically-thick regime. We therefore conjecture that this orphan $K_s$-band flare is produced by optically-thick synchrotron emission from an outer jet. 
If the optically-thick synchrotron emission extends up to the $R$-band with a flat spectral shape and the baseline $R$-band flux is not as high as the flaring component of $\sim 2.0$~Jy (after extinction correction, assuming $A_V=4.0$), it significantly contributes to the $R$-band flux as well, making the flare visible also in the $R$-band light curve. 
A spectral break between the $K_s$ and $R$ bands of the flaring emission component caused by the transition of synchrotron emission from optically-thick to optically-thin regimes, if present, would make the contribution of the flaring component negligible with respect to the baseline flux, as observed. 
The quasi-simultaneous $F_{\nu}$ \kanata/HONIR and \swift/XRT spectra measured on MJD~57194.52, together with the RATAN-600 non-simultaneous radio fluxes are also shown. The radio, optical/NIR, and X-ray spectra would be reasonably understood as optically-thick synchrotron emission from the outer jet, disk emission, and disk plus corona emissions, respectively.

\begin{figure*}[!th] 
\begin{center}
\plotone{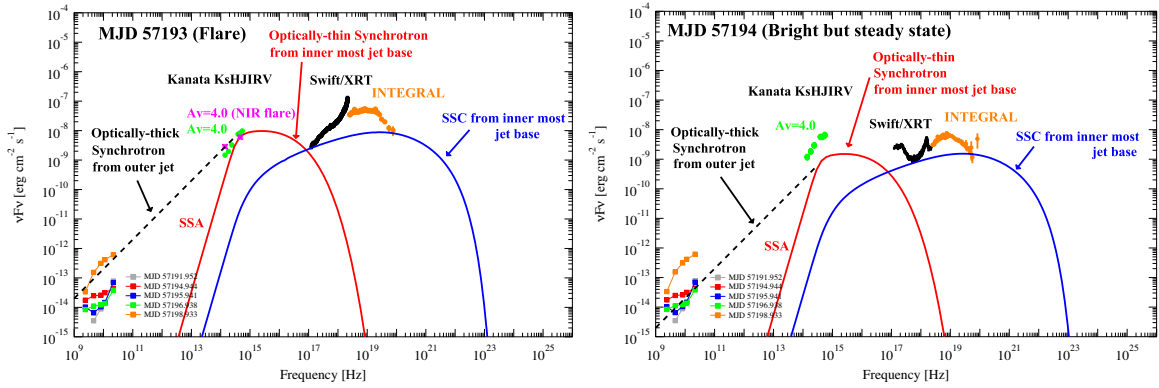}
\caption{{\it Left}: Broadband spectrum of \src\ in $\nu - \nu F_{\nu}$ representation. The \kanata/HONIR and \swift/XRT fluxes were quasi-simultaneously observed on MJD~57193, while RATAN-600 and \integral\ data were not simultaneous. 
One-zone synchrotron and SSC emissions are also drawn by solid red and blue lines, respectively. Black dashed line indicates the optically-thick synchrotron emission from an outer jet, with a spectral shape of $F_{\nu} \propto \nu^0$.
{\it Right}: Same as the {\it left} panel, but for the quasi-simultaneous \kanata/HONIR and \swift/XRT fluxes observed on MJD~57194.
}
\label{fig:nufnu}
\end{center}
\end{figure*}

\subsection{Properties of the jet}
\label{sec:jet}
To constrain the jet parameters and physical quantities in the emission region, we attempted to model the spectral energy distribution (SED) of \src\ by using a one-zone synchrotron plus synchrotron self-Compton (SSC) model \citep{Finke08}, which is widely used for blazar SED modeling \citep[e.g.,][]{Abdo09, Abdo11, Tanaka14, Tanaka15}.
We show in Fig.~\ref{fig:nufnu} ({\it left} panel) the quasi-simultaneous broadband SED of \src\ during the $K_s$-band flare.
This modeling assumes that a single emission region is located at the inner-most part of the jet. Hence, the optically-thick synchrotron emission from an outer jet, which has a flat spectrum of $F_{\nu} \propto \nu^0$ observed in the radio up to NIR band is not modeled, while the optically-thin synchrotron and SSC emissions at optical frequencies and higher are fitted. However, in the current case, we now know that the optical and X-ray emission are from a disk and disk plus corona, respectively. We therefore regard the \kanata/HONIR, \swift/XRT, and \integral\ data points \citep[taken from Fig.~3 of][]{Rodriguez15} as upper limits for the jet emission. Another constraint comes from the \kanata/HONIR observation that the $K_s$-band emission is not significantly polarized even during the orphan flare. This indicates that the $K_s$-band emission is still in the optically-thick regime and that the break frequency (defined as $\nu_{\rm SSA}$) is due to synchrotron self absorption (SSA). The transition from the optically-thick to optically-thin regime is above the $K_s$-frequency band, thus $\nu_{\rm SSA} \gtrsim 1.4 \times 10^{14}$~Hz, and we adopt a value of $\nu_{\rm SSA}=3.0\times 10^{14}$~Hz. We also assume that the synchrotron peak flux is 2~Jy as observed by \kanata/HONIR (and also by RATAN-600 on MJD~57198.933). We can then derive the magnetic field $B$ and the size of the emission region $R$ by using the standard formulae for synchrotron absorption coefficient and emissitivity \citep[e.g.,][]{Rybicki, Chaty11, Shidatsu11},
\begin{eqnarray}
B &\approx& 1 \times 10^5 \left( \frac{\nu_{\rm SSA}}{3 \times 10^{14} \ {\rm Hz}} \right)^{\frac{3s+10}{2s+13}} \left( \frac{F_{\nu}}{2 \ {\rm Jy}} \right)^{-\frac{2}{2s+13}} \left( \frac{D}{2.4 \ {\rm kpc}} \right)^{-\frac{4}{2s+13}} \ {\rm Gauss}, \label{eq:mag}\\
R &\approx& 5 \times 10^8 \left( \frac{\nu_{\rm SSA}}{3 \times 10^{14} \ {\rm Hz}} \right)^{\frac{-(s^2+7s+8)}{2(2s+13)}} \left( \frac{F_{\nu}}{2 \ {\rm Jy}} \right)^{\frac{s+6}{2s+13}} \left( \frac{D}{2.4 \ {\rm kpc}} \right)^{\frac{2(s+6)}{2s+13}} \ {\rm cm}, \label{eq:size}
\end{eqnarray}
where $s$ is the power-law index of electron distribution (see below) and $D$ is the distance to \src. Here we assumed almost equipartition between magnetic field and electron energy density, which was confirmed by the following SED modeling.

The electron energy distribution is assumed to have a single power-law shape with exponential cutoff as $dN/d\gamma = K \gamma^{-s} \exp(- \gamma/ \gamma_{\rm cut})$ for $\gamma_{\rm min} \leq \gamma \leq \gamma_{\rm max}$, where $\gamma$ is the electron Lorentz factor, $K$ is the electron normalization, $s$ is the power-law index, $\gamma_{\rm cut}$ is the cutoff energy, $\gamma_{\rm min}$ and $\gamma_{\rm max}$ are the minimum and maximum electron energies and are respectively set to 1 and $10^6$. The jet inclination angle of \src\ is estimated as $\sim 55^{\circ}$ \citep[e.g.,][]{Shahbaz94, Khar10} and assuming a jet velocity of $0.9c$, the corresponding Doppler beaming factor is $\delta = \left[ \Gamma \left(1- \beta \cos \theta \right) \right]^{-1} \sim 0.9$. We can therefore safely neglect relativistic beaming effects. 

By changing the parameters of the electron energy distribution, we calculated the resultant synchrotron and SSC emissions. The calculated model curves are shown in Fig.~\ref{fig:nufnu} ({\it left} panel) and all the model parameters are tabulated in Table~\ref{tab:model}. The derived parameters for the electron energy distribution are $K=4.5\times 10^{40}$, $s=2.2$, and $\gamma_{\rm cut}=10^2$. Importantly, the \swift/XRT data allowed us to constrain the cutoff energy as $10^2$ because larger $\gamma_{\rm cut}$ violates these upper limits in the soft X-ray band. This implies that particle acceleration in this microquasar jet is not very efficient. The electron energy distribution cutoff energy is determined by the balance between acceleration and cooling times, where the acceleration time is defined as $t_{\rm acc}= \eta E/\left(eBc \right)$ by using an electron energy $E$ and parameter $\eta$, the number of gyrations an electron makes while doubling its energy \citep[e.g.,][]{Finke08, Murase14}. Since dominant cooling processes for electrons of $\gamma=\gamma_{\rm cut}=10^2$ are both synchrotron and SSC, the cooling time is estimated as $t_{\rm cool}=3 \gamma m_e c/ \left( 8 \sigma_T \gamma^2 \left(U_{\rm B}+U_{\rm sync} \right) \right)$, where $U_{\rm B}=B^2/8 \pi$ and $U_{\rm sync}=L_{\rm sync}/4 \pi R^2 c$ are the energy densities of magnetic field and synchrotron photons, respectively \citep[e.g.,][]{Finke08}. Thus, we obtain $\eta \sim 10^6$ by setting $E=\gamma_{\rm cut} m_e c^2$ and $B=10^5$~G. This is much larger than $\eta \sim 10$ in blazar jets \citep[e.g.,][]{Rachen98}, indicating much longer acceleration times and inefficient acceleration in this microquasar jet.

Note that the electron power-law index of 2.2 we obtained from SED modeling has already been modified by rapid synchrotron and SSC cooling.
In the current situation, because the magnetic field is strong ($B \sim 1 \times 10^5$~G) and emission region is small ($R \sim 5 \times 10^8$~cm), we need to consider the following three energy loss processes: adiabatic cooling (this is also equivalent to particle escape from emission region), synchrotron cooling, and SSC cooling. Note here that we can neglect synchrotron cooling for electrons of $\gamma \lesssim 40$ because the optically-thick regime is below $\nu_{\rm SSA}=3 \times 10^{14} \left( \gamma/40 \right)^2 \left( B/10^5 \ {\rm G} \right) \ {\rm Hz} $ \citep[e.g.,][]{Piran04}. Cooling timescales for these processes are estimated as $t_{\rm ad} \gtrsim R/c \sim 1.7\times10^{-2} \left( R/5\times10^8 \ {\rm cm} \right)$ and $t_{\rm SSC} \sim 5.1 \times 10^{-2} \gamma^{-1} \left( U_{\rm sync} / 3 \times 10^8 \ {\rm erg} \ {\rm cm}^{-3} \right)^{-1}$, respectively \citep[e.g.,][]{Piran04, Chaty11}. Therefore, high-energy electrons of $\gamma \gtrsim 3$ rapidly lose their energy via SSC emission and hence the electron power-law index becomes steeper by one power of $E$, if injection of high-energy emitting electrons continued over a few tens of minutes (which corresponds to the flare duration observed by \kanata/HONIR). Namely, the electron energy distribution at $\gamma \gtrsim 3$ is already in a fast-cooling regime, which indicates that the original (or injected) electron power-law index is 1.2. This is much smaller than the standard power-law index of 2.0 derived by the first-order Fermi acceleration theory \citep[e.g.,][]{Blandford78}.

We can now derive the total energy in electrons and magnetic field as $W_{\rm e}=m_e c^2 \int_{\gamma_{\rm min}}^{\gamma_{\rm max}} d\gamma \gamma dN/d\gamma = 9.9 \times 10^{34}$ erg and $W_{\rm B}=\left(4/3\right) \pi R^3 U_{\rm B} = 4.7\times 10^{35}$~erg, respectively, thus the jet is Poynting-flux dominated by a factor of $\sim 5$. The jet power in electrons ($L_{\rm e}$) and magnetic field ($L_{\rm B}$) is calculated as $L_{\rm i}=2 \pi R^2 \beta c U_{\rm i} \ \left( {\rm i=e, B}\right)$, where $U_{\rm e}=W_{\rm e}/ \left( 4 \pi R^3/3 \right)$ is the electron energy density, $\beta=0.9$ is assumed, and the factor of 2 is due to the assumption of a two-sided jet \citep[e.g.,][]{Finke08}. Then, we obtain $L_{\rm e}=7.8 \times 10^{36}$ erg s$^{-1}$ and $L_{\rm B}=3.7 \times 10^{37}$ erg s$^{-1}$, and the summed power ($L_{\rm e}+L_{\rm B}$) amounts to $4.5 \times 10^{37}$ erg s$^{-1}$. On the other hand, we can also calculate the total radiated power using the SED modeling result as $L_{\rm rad}=L_{\rm sync}+L_{\rm SSC}= 7.0 \times 10^{37}$ erg s$^{-1}$, which is larger than the summed $L_{\rm e}+L_{\rm B}=4.5 \times 10^{37}$ erg s$^{-1}$. This indicates that the Poynting flux ($L_{\rm B}$) and $L_{\rm e}$ are not sufficient to explain $L_{\rm rad}$ and another form of power is required. The simplest and most probable solution is to assume that the jet contains enough protons which have larger power than $L_{\rm rad}$ \citep[e.g.,][]{Sikora00, Ghisellini14, Tanaka15, Saito15}. This is another (though indirect) evidence of a baryon component in a microquasar jet. Note that we reached the above conclusion based on jet energetics argument, but the baryonic jet in a microquasar has already been claimed by a different, independent method based on the detection of blue-shifted emission lines in the X-ray spectra for SS~433 and 4U~1630$-$47 \citep{Kotani94, Trigo13}.

By assuming that the jet contains one cold proton per one relativistic (emitting) electron, we can derive the total energy of cold protons as $W_{\rm p}=m_p c^2 \int_{\gamma_{\rm min}}^{\gamma_{\rm max}} d\gamma dN/d\gamma = 5.4 \times 10^{37}$ erg, where $m_p$ is the proton mass. This corresponds to the cold proton power ($L_{\rm p}$) of $2.1 \times 10^{39}$ erg s$^{-1}$ by using the relation of $L_{\rm p}=2 \pi R^2 \beta c U_{\rm p}$ \citep[e.g.,][]{Finke08}, where $U_{\rm p}=W_{\rm p}/ \left( 4 \pi R^3/3 \right)$ is the energy density of cold protons and $\beta=0.9$ is assumed. We therefore obtain the total jet power $L_{\rm jet}=L_{\rm e}+L_{\rm B}+L_{\rm p}$ and radiative efficiency of the jet as $L_{\rm rad}/L_{\rm jet} \sim 3\%$ (see also Table~\ref{tab:model}). 

During the moderate and steady state on MJD~57194, there were no observational constraints on $\nu_{\rm SSA}$ due to the dominance of the disk component in the optical and NIR bands, hence we assume that {\it it remained the same as that during the bright flare,} $\nu_{\rm SSA}=3.0\times 10^{14}$~Hz. We estimated the synchrotron peak flux as 0.2 Jy, because such a flux level was observed in the GHz band during the high state \citep{RATAN_ATel} and an extrapolation to the NIR band with a flat shape ($\alpha=0$) seems reasonable. We thereby obtained the following estimates of $B \sim 2 \times 10^5$~G and $R \sim 2 \times 10^8$~cm (see Equations~(\ref{eq:mag}) and (\ref{eq:size})). Important information about the non-thermal jet emission, which should be included in the SED modeling, comes from the \integral\ detection of an additional power-law component of $\Gamma=1.54^{+0.24}_{-0.45}$ in the hard X-ray band \citep{Rodriguez15}. The \kanata/HONIR, \swift/XRT, and \integral\ (exponential cutoff power-law component dominant up to $\sim 100$~keV) data points are treated as upper limits. Based on these assumptions and the multi-wavelength data, we calculated the broadband non-thermal jet emission by accelerated electrons using the one-zone synchrotron and SSC model. The result is shown by solid lines in Fig.~\ref{fig:nufnu} ({\it right} panel) and the model parameters are tabulated in Table~\ref{tab:model}. We obtained the same parameter values of $s=2.2$ and $\gamma_{\rm cut}=10^2$ for the electron energy distribution, but the electron normalization $K=7.0 \times 10^{39}$ is smaller due to the fainter jet flux, as was observed in the radio band. We also found $L_{\rm B}/L_{\rm e} \sim 2$, suggesting again the jet is slightly Poynting-flux dominated. More interestingly, the total radiated power is again larger than the summed electron and magnetic field powers in the jet (see Table~\ref{tab:model}), implying the presence of a baryonic component even during the fainter state.

Our discussion of the jet properties of \src\ in this section, particularly in relation to AGN jets can be summarized as follows.
\begin{enumerate}
\item The SSA frequency and peak flux density enable us to estimate the magnetic field strength and size of the emission region by assuming equipartition between magnetic field and relativistic electrons. The derived magnetic field of $B \sim 10^5$~Gauss is much stronger, and size of the emission region of $R\sim10^8$~cm much smaller, compared to AGN jets (typically $B \sim 1-10$~Gauss and $R \sim 10^{17-18}$~cm, see e.g., \citet{Ghisellini10}).
\item Based on modeling of the broadband spectrum of \src, we found an upper limit to the cutoff Lorentz factor of electrons of $\sim 100$. Because the cutoff is determined by the balance of the acceleration and cooling times, this result  implies a longer acceleration time of $\eta \sim 10^6$ in this microquasar jet, suggesting electron acceleration is much less efficient compared to AGN jets (that typically show $\eta \sim 10$).
\item The original (or injected) power-law index of the electron energy distribution was derived as $s=1.2$. In the SED modeling of AGN jets, electrons are assumed to have a broken power-law shape. The power-law index below the break Lorentz factor $\gamma_{\rm break}$ (typically $\gamma_{\rm break} \sim 100-1000$) is estimated as $s \sim 1$ \citep[e.g.,][]{Ghisellini10}. Hence, $s=1.2$ derived here in \src\ jet is comparable to that derived in AGN jets, implying that same acceleration mechanism operates in these different systems.
\item To account for the total radiated power of the jet of \src, a cold proton component is required inside the jet. This is the same situation as in AGN jets.
\item During the bright flare on MJD~57193, the jet radiative efficiency ($\epsilon_{\rm rad} \equiv L_{\rm rad}/L_{\rm jet} = L_{\rm rad}/\left( L_{\rm e}+L_{\rm B}+L_{\rm p} \right)$) is derived as $\sim 3\%$. This is roughly comparable to that of AGN and GRB jets ($\sim 10\%$, see e.g., \citet{Nemmen12, Ghisellini14}).
\end{enumerate}


\acknowledgments
We appreciate the anonymous referee's constructive comments that helped to improve the manuscript.
We thank Dr.~Trushkin and Dr.~Rodriguez for providing us with their RATAN-600 and \integral\ data, respectively.
This work is supported by the Optical \& Near-Infrared Astronomy Inter-University Cooperation Program, and the MEXT of Japan.
We acknowledge with thanks the variable star observations from the AAVSO International Database contributed by observers worldwide and used in this research. 
YTT is supported by Kakenhi 15K17652. MU is supported by Kakenhi 25120007. CCC is supported at NRL by NASA DPR S-15633-Y. MW is supported by Kakenhi 25707007. 

\bibliographystyle{apj}

\begin{table}
\scriptsize
\begin{center}
\caption{Daily optical and near-infrared polarization degree (PD) and position angle (PA).}
\label{tab:pol}
\begin{tabular}{lccccc}
\hline
 & \multicolumn{2}{c}{$R$-band}  &  \multicolumn{2}{c}{$K_s$-band}  & \\
      MJD  & PD (\%)& PA (deg) & PD (\%) & PA (deg) & Instrument \\
\hline
57190 & $7.85\pm0.11$ (1.34/25) & $6.59\pm0.35$ (1.09/25) & -- & -- & \pirka/MSI \\
57191 & $7.72\pm0.05$ (0.62/20) & $6.33\pm0.22$ (1.04/20) & -- & -- & \pirka/MSI \\
57192 & $7.76\pm0.02$ (0.51/22) & $6.39\pm0.10$ (0.74/22) & -- & -- & \pirka/MSI \\
57193 & $7.77\pm0.01$ (0.74/226) & $6.19\pm0.03$ (0.72/226) & -- & -- & \pirka/MSI \\
           & $8.03\pm0.06$ (0.24/16) & $8.23\pm0.25$ (0.29/16) & $1.43\pm0.10$ (1.45/14) & $9.13\pm2.18$ (1.73/14) & \kanata/HONIR \\
57194 & $7.96\pm0.02$ (0.58/44)  & $7.92\pm0.07$ (0.32/44) & $1.50\pm0.04$ (0.78/24) & $10.36\pm0.72$ (0.69/24) & \kanata/HONIR \\
\hline
\end{tabular}
\tablecomments{Shown in the parentheses are corresponding reduced $\chi^2$ and degree of freedom when the observations were fitted with a constant value (i.e., assuming no temporal variation).}
\end{center}
\end{table}

\begin{table}
\footnotesize
\begin{center}
\caption{Model parameters.}
\label{tab:model}
\begin{tabular}{lccc}
\hline
Parameter & Symbol & MJD 57193 & MJD 57194 \\
\hline
Break frequency [Hz] & $\nu_{\rm SSA}$	& $3 \times 10^{14}$ & $3 \times 10^{14}$ \\
Magnetic Field [G] & $B$         & $1.4 \times 10^5$ & $1.8 \times 10^5$ \\
Size of emission region [cm] & $R $ & $5.3 \times 10^8$ & $1.7 \times 10^8$ \\
Jet velocity [c] & $\beta$ & $0.9$ & $0.9$ \\
\hline
Electron distribution normalization [electrons] & $K$ & $4.5 \times 10^{40}$ & $7.0 \times 10^{39}$ \\
Electron Power-law Index & $s$   & 2.2 & 2.2 \\
Minimum Electron Lorentz Factor & $\gamma_{\rm min}$  & $1.0$ & $1.0$ \\
Cutoff Electron Lorentz Factor & $\gamma_{\rm cut}$ & $10^2$ & $10^2$ \\
Maximum Electron Lorentz Factor & $\gamma_{\rm max}$  & $10^6$ & $10^6$ \\
\hline
Synchrotron luminosity [$\erg\ \s^{-1}$] & $L_{\rm sync}$ & $2.8\times10^{37}$ & $4.3\times10^{36}$ \\
SSC luminosity [$\erg\ \s^{-1}$] & $L_{\rm SSC}$ & $4.1\times10^{37}$ & $7.2\times10^{36}$ \\
Total radiation luminosity [$\erg\  \s^{-1}$] & $L_{\rm rad}$ & $6.9\times10^{37}$ & $1.2\times10^{37}$ \\
\hline
Jet Power in Magnetic Field [$\erg\ \s^{-1}$] & $L_{\rm B}$ & $3.7\times10^{37}$ & $6.3\times10^{36}$ \\
Jet Power in Electrons [$\erg\ \s^{-1}$] & $L_{\rm e}$ & $7.8\times10^{36}$ & $3.7\times10^{36}$ \\
Jet Power in Cold Protons [$\erg\ \s^{-1}$] & $L_{\rm p}$ & $2.1\times10^{39}$ & $2.0\times10^{39}$ \\
\hline
Jet Radiative Efficiency [\%] & $\epsilon_{\rm rad}$ & $\sim 3$ & $\sim 1$ \\
\hline
\end{tabular}
\end{center}
\end{table}

\end{document}